\documentclass[prl,twocolumn,superscriptaddress,showpacs]{revtex4}
\usepackage{amsmath}
\usepackage{amsthm}
\usepackage{amssymb}
\usepackage{amsbsy}
\usepackage{graphicx}

\begin{document}

\newcommand{\smallrech}{ \; \pspicture[0.35](0.2,0.1)
\psset{linewidth=0.025,linestyle=solid}
\psline[](0.1,0)(0.1,0.1)
\pspolygon[](0,0)(0.2,0)(0.2,0.1)(0,0.1)
\endpspicture\;}
\newcommand{\orients}{ \; \pspicture[0.35](0.01,0.01) \psarc{->}(-0.01,0.12){0.1}{10}{350} \endpspicture\;}

\newcommand{\smallrecv}{ \; \pspicture[0.35](0.1,0.2)
\psset{linewidth=0.025,linestyle=solid}
\psline[](0,0.1)(0.1,0.1)
\pspolygon[](0,0)(0,0.2)(0.1,0.2)(0.1,0.0)
\endpspicture\;}

\newcommand{\emptya}{  \; \pspicture[0.35](0.6,0.2)
\psdots[linecolor=black,dotsize=.10]  (0.15,0)(0.15,0.3)  (0.6,0.15)
\psset{linewidth=0.03,linestyle=dotted}
\psline[](0.3,0)(0.3,0.3)
\pspolygon[](0,0)(0.6,0)(0.6,0.3)(0,0.3)
\psset{linewidth=0.05,linestyle=solid}
\psline[linecolor=black](0,0)(0.3,0)
\psline[linecolor=black](0,0.3)(0.3,0.3)
\psline[linecolor=black](0.6,0)(0.6,0.3)
\endpspicture \;}

\newcommand{\emptyb}{ \; \pspicture[0.35](0.6,0.2)
\psdots[linecolor=black,dotsize=.10]  (0.45,0)(0.45,0.3)(0,0.15)
\psset{linewidth=0.03,linestyle=dotted}
\psline[](0.3,0)(0.3,0.3)
\pspolygon[](0,0)(0.6,0)(0.6,0.3)(0,0.3)
\psset{linewidth=0.05,linestyle=solid}
\psline[linecolor=black](0.3,0)(0.6,0)
\psline[linecolor=black](0.3,0.3)(0.6,0.3)
\psline[linecolor=black](0,0)(0,0.3)
\endpspicture \;}

\newcommand{\fulla}{ \; \pspicture[0.35](0.6,0.2)
\psdots[linecolor=black,dotsize=.10]   (0.15,0)(0.15,0.3)  (0.6,0.15)(0.3,0.15)
\psset{linewidth=0.03,linestyle=dotted}
\psline[](0.3,0)(0.3,0.3)
\pspolygon[](0,0)(0.6,0)(0.6,0.3)(0,0.3)
\psset{linewidth=0.05,linestyle=solid}
\psline[linecolor=black](0,0)(0.3,0)(0.3,0.3)(0,0.3)
\psline[linecolor=black](0.6,0)(0.6,0.3)
\endpspicture\;}

\newcommand{\fullb}{ \; \pspicture[0.35](0.6,0.2)
\psdots[linecolor=black,dotsize=.10]  (0.45,0)(0.45,0.3)  (0,0.15)(0.3,0.15)
\psset{linewidth=0.03,linestyle=dotted}
\psline[](0.3,0)(0.3,0.3)
\pspolygon[](0,0)(0.6,0)(0.6,0.3)(0,0.3)
\psset{linewidth=0.05,linestyle=solid}
\psline[linecolor=black](0.6,0)(0.3,0)(0.3,0.3)(0.6,0.3)
\psline[linecolor=black](0,0)(0,0.3)
\endpspicture \;}

\bibliographystyle{apsrev}

\title{First- order versus unconventional phase transitions in three-dimensional dimer models}

\author{Stefanos Papanikolaou}
\affiliation{Laboratory of Atomic and Solid State Physics, Cornell University, Ithaca, New York 14853-2501, USA}

\author{Joseph J. Betouras}
\affiliation{Scottish Universities Physics Alliance,School of Physics and Astronomy,
University of St Andrews, North Haugh, St Andrews, Fife KY16 9SS, UK.}

\date{\today}

\begin{abstract}

We study the phase transition between the Coulomb liquid and the columnar crystal in the 3D classical dimer model, which was found to be continuous in the O(3) universality class~\cite{alet1}. In addition to nearest neighbor interactions which favor parallel dimers, further neighbor interactions are allowed in such a manner that the cubic symmetry of the original system remains intact. We show that the transition in the presence of weak additional, symmetry preserving interactions is first-order. However the universality class of the transition remains continuous when the additional interactions are weakly repulsive. In this way, we verify the existence of a multicritical point near the unperturbed transition and we identify a critical line of unconventional transitions between the Coulomb liquid phase and the $6-$fold columnar phase.
%Thus, the unperturbed transition consists of a fine - tuned multicritical point and in the light of these results,  we discuss the consequences for the available scenarios~\cite{powell, charrier} that attempted to explain the nature of the transition in the absence of the additional interactions. 
\end{abstract}

\pacs{xxxx}
\maketitle

Phase transitions are fundamental in nature and as a consequence they are relevant to every branch of classical or quantum physics. They also serve as an important  playground for conceptual development of new ideas. A powerful approach to phase transitions and critical phenomena is the Landau-Ginzburg-Wilson (LGW) theory~\cite{wilson} which is based on the concept of a local order parameter which characterizes a phase where some symmetry is spontaneously broken. Then, the corresponding free energy of the system in  the region around the critical point is expanded in powers of the order parameter and several properties can be studied using renormalization group methods for continuous phase transitions. Recently, Senthil \emph{et al}. \cite{senthil} proposed that phase transitions between phases with different non-trivial symmetries can be, in principle, continuous without requiring fine-tuning. In particular, in the context of quantum antiferromagnets, it is thought that such a continuous phase transition from the Neel state to a valence bond solid can be realized~\cite{sandvik}, contradicting the LGW prediction of a first order transition and its recent observation in recent simulations~\cite{sirker}, an intermediate disordered phase or a \emph{fine-tuned} direct continuous transition. 

In particular, phase transitions in systems composed of classical dimers at close packing could serve as an important example to illustrate the idea of a non-LGW thermal transition~\cite{diep}. In such lattice systems, links are covered by close-packed hard-core dimers. The constraint of having precisely one dimer emanating from each site of the  lattice allows for plausible connections to be made with associated gauge theories~\cite{kogut},  capturing the essence of non-LGW transitions. In the case of the three dimensional (3D) cubic classical dimer model, the development of a featureless liquid with short-range correlations is not allowed even at infinite temperatures because then the dimer correlations fall off algebraically with the distance, being dipolar, and gapless collective modes (photons) are present~\cite{huse}. This so-called \emph{Coulomb phase} has also been found in other three dimensional models~\cite{villain} and is relevant to some pyrochlore compounds~\cite{isakov, henley}.
In the presence of local attractive interactions, strong evidence was put forth~\cite{alet1, misguich} that there is a continuous transition from the Coulomb liquid phase to a 6-fold degenerate columnar phase in the O$(3)$ universality class, which could support a non - LGW theoretical description. However, the question still remains open whether the transition is highly fine-tuned or weakly first-order. Several theoretical scenarios have been suggested for the nature of this transition~\cite{powell, charrier}, proposing that the transition has an effective $SU(2)$ symmetry, and some evidence is linked to a mapping on a 2D quantum problem~\cite{powell}. It is argued that this symmetry is not broken, since high-order terms are required to violate the symmetry. Here, we study the same problem in the presence of weak, same symmetry but apparently higher order perturbations, seemingly irrelevant in the RG sense.  

The system that we are interested in~\cite{alet1, huse} is a cubic lattice of volume $V=L^3$ where L is the linear system size, covered by  hard-core dimers and interactions which favor parallel alignment of dimers and respect the cubic symmetry. The energy is determined by the number of parallel dimers in x, y and z directions at different distances but same column $E = E_{=} + E_{||} + E_{//}$ with

\begin{eqnarray}
E_{\alpha}=-J_1\sum_{{\bf r}\in {\rm links}}\left[ N^{(1)}_{\alpha}({\bf r})+ \frac{J_2}{J_1}N^{(2)}_{\alpha}({\bf r})+  \frac{J_3}{J_1}N^{(3)}_{\alpha}({\bf r})\right]\; ,
\end{eqnarray}
\begin{equation}
N_{\alpha}^{(i)}({\bf r})= n_{m_\alpha}({\bf r })n_{m_\alpha}({\bf r}+i\; \hat e_{\alpha})\; ,
\end{equation}
where $\alpha $ can be $=,{||},{//}$, i.e. pair of parallel dimers along the x, y or z directions and correspondingly $m_\alpha = x, y, z$. By $n_{\hat{m}}(i)$ we denote the absence or existence (0 or 1 respectively) of dimer on the link ($\vec{r}_i, \vec{r}_i+\hat{e}_m$) of the lattice, where $\hat{e}_m$ is the unit vector of the lattice in each of the three directions ($m=x,y,z$). Note that the dimer orientation is perpendicular to the direction along which the interactions are considered.  

\begin{figure}[t]
\includegraphics[width=0.7\columnwidth]{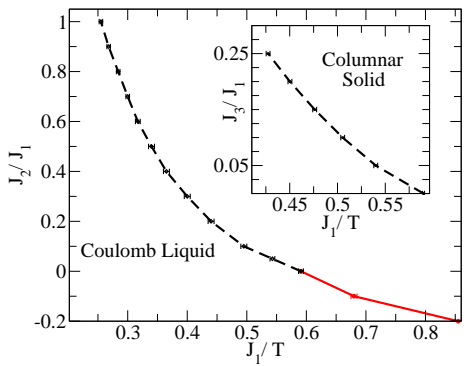}
\centering
\caption{\label{fig:1}{\bf  Phase diagrams of the described models in the presence of the interaction $J_2$ (and $J_3=0$) (inset: $J_3$, with $J_2=0$)}. In both cases, the lines shown separate the Coulomb liquid phase from the columnar phase.The point where $J_2$($J_3$) is zero is consistent with the result reported in Ref.~\cite{alet1}.  The dashed lines denote first-order transitions whereas the solid signify continuous ones. The lines through the points are guide to the eye.}
\end{figure}

The common property of the perturbations $J_2, J_3$ is that they have the same sign with the original interaction, and they respect the cubic symmetry. We choose both $J_2$ and $J_3$ to be attractive, since if repulsive, there is additional frustration, similar to that of the ANNNI model~\cite{fisher, papanikolaou2}, the study of which is the subject of future work~\cite{spjb}. 
Given the order parameter of the transition,
\begin{equation}
m_{\alpha}({\bf r})=\frac{1}{2}(-1)^{r_\alpha}[n_\alpha({\bf r})-n_\alpha({\bf r}-{\hat e}_\alpha)] \; ,
\end{equation}
the two sets of ground-states, classified according to their order parameter expectation value, favored by $J_2$ and $J_3$ respectively, are essenially different.  

 In this paper, we show that (I) Small perturbations that keep the original cubic symmetry of the Hamiltonian intact and have the same sign (attractive), drive the transition from the Coulomb liquid to the columnar solid first - order, (II) The first-order transition gets stronger as the strength of the perturbation increases, (III) Our study is consistent with having a continuous transition for the system without the interactions $J_2$, $J_3$, or when they are repulsive.

We perform calculations with lattices of linear size L=12, 16, 20, 24 and periodic boundary conditions, using the Wang-Landau Monte Carlo algorithm~\cite{wanglandau} which is efficient in detecting first order transitions. The Wang-Landau scheme determines the density of states g(E) independently of the temperature, therefore there is a direct access to the Gibbs probability histograms.
Independent random walks are performed in different, restricted ranges of energy, by flipping plaquettes occupied by nearest neighboring parallel dimers. 
 
\begin{figure}[t]
\includegraphics[width=0.75\columnwidth]{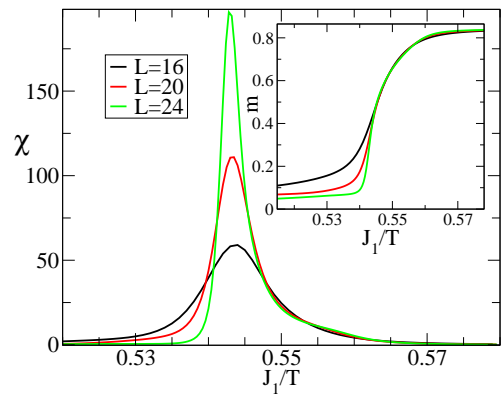}
\centering
\caption{ {\bf The order parameter $m$ (inset) and its susceptibility for $J_2/J_1=0.05$}. The visible "shoulder" 
feature in the specific heat does not appear here. The peak of the susceptibility, which signifies that the high coupling phase is columnar,
 scales as $L^3$ as one expects for a first-order transition.}
\label{fig:2}
\end{figure}

Each time a state of energy E is visited, we update the corresponding density of states by multiplying the existing value by a
modification factor $f >1$, $g(E) \rightarrow g(E) f$. The density of
states is modified until the accumulated histogram $H(E)$ is relatively flat and it converges to the true value
with an accuracy proportion to $\log(f)$. By refining continuously the modification factor ($f\rightarrow f/\sqrt{2}$) and repeating the prescribed procedure, we achieve $10^{-6}$ accuracy for the density of states.

\begin{figure}[t]
\includegraphics[width=0.75\columnwidth]{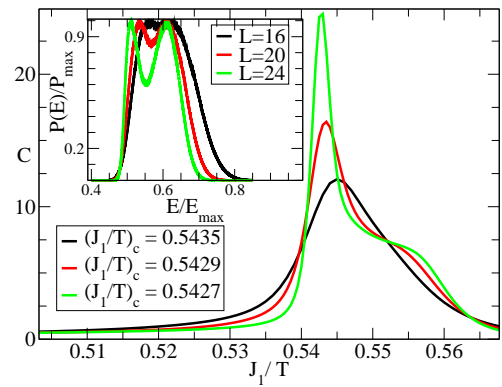}
\caption{{\bf Specific heat of the model with $J_2=0.05J_1$}. The peak scales asymptotically as $\sim L^3$, expected for a first-order transition~\cite{challa}. The "shoulder" feature, observed in Ref.~[\onlinecite{alet1}] is also present here, even though much weaker and goes away as the interaction strength gets bigger. In the inset, the energy histograms develop a 2-peak structure for $J_3=0.05J_1$, characteristic of a first-order transition. }
\label{fig:3}
\end{figure}

\begin{figure}[t]
\includegraphics[width=0.75\columnwidth]{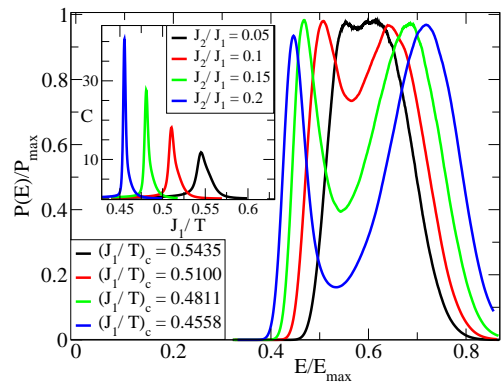}
\centering
\caption{{\bf First-Order transition signature}. The distance between the two peaks in the energy histograms gets bigger, as the interactions ($J_2$) gets larger, as it is shown for $L=16$. In the inset, the specific heat develops a larger peak (for the same lattice size $L=16$) as the interaction gets larger. A very similar picture holds also for the case of attractve $J_3$ interactions.}
\label{fig:4} 
\end{figure}

\begin{figure}[t]
\includegraphics[width=0.75\columnwidth]{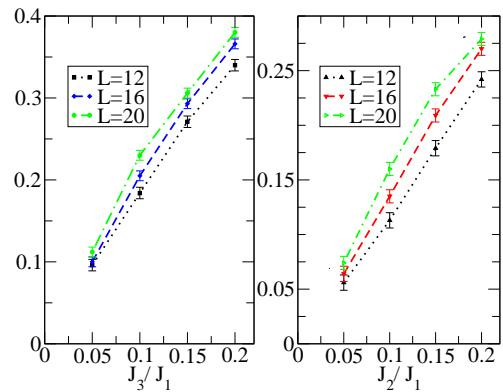}
\centering
\caption{ {\bf Weakness of the first-order transition for small perturbations}. The distance between the two energy probability peaks when they have equal height, as a function of the interaction strength, for the two types of interactions we studied. Clearly, our study is consistent with a continuous transition for the parent system, since the slope becomes sharper near $(0,0)$, making a non-zero value at $J_{2,3}/J_1=0$ improbable. The dashed lines are guides to the eye.
}
\label{fig:5}
\end{figure}

The phase diagrams of the 3D dimer model in the presence of $J_2$ or $J_3$ are shown in Fig.~\ref{fig:1}.
The transition temperatures are specified by using the characteristic features that the energy cumulant $V_L=1-\frac{\langle E^4\rangle}{3\langle E^2\rangle^2}$ has at a first order transition~\cite{challa}. More specifically, $V_L$ has a characteristic dip at the first-order transition point and takes the value $2/3$ at all other couplings. The position of this dip converges exponentially fast with the system size and extrapolation gives an accurate estimate of $(J_1/T)_c$.
Our results are consistent, in the absence of any additional perturbations, with the finding by Alet \emph{et al.} of $J_{1 c}=0.597$~\cite{alet1}. As the interactions grow though, the transition becomes first order, with a decreasing correlation length. The transition moves to smaller couplings, as expected since the additional interactions are attractive. For the smallest coupling available $J_2/J_1=0.05$, the specific heat, as it is shown in Fig.~\ref{fig:3}, shows the big "shoulder" feature that was observed by Alet \emph{et al.} in the parent system, but is remarkably weaker, and gradually vanishes as the interaction gets stronger. As it is shown in Fig.~\ref{fig:4}, the characteristic 2-peak structure for the energy probability histograms, gets widened as the interaction grows and at the same time the specific heat peak gets larger, for the same lattice size. Also, by tracking the distance between the peaks of the energy histograms as a function of the interaction strength $J_2/J_1$ or $J_3/J_1$, we found that there is a consistency of our result with having a continuous transition in the absence of any additional interactions, as it was found by Alet~\emph{et al.}~\cite{alet1}. The distance between the peaks gets closer to zero, if extrapolated to zero coupling $J_2$ or $J_3$, as shown in Fig.~\ref{fig:5}. 

We also studied the case where the interactions $J_2$ are repulsive. We find, by performing simulations, that for small $J_2/J_1=-0.1,-0.2,-0.4$ there is no signature of double - peak structure in the histogram structure, showing no signs of a discontinuous transition~(cf. Supplementary material). Moreover, the scaling behavior of the susceptibility and specific heat peaks, signify that the transition has exponents that are close to the ones found in Ref.~\cite{alet1}, ($\alpha/\nu=0.75-1.0$ and $\gamma/\nu=1.75-2.0$, giving $\nu=0.5-0.53$ if we use hyperscaling) even though our simulated system sizes cannot permit us to conclude whether the transition for $J_2/J_1<0$ is in the same universality class.

In Ref. \cite{powell} it was suggested that the critical theory of the classical dimer problem on the cubic lattice with nearest neighbor interactions is SU(2) invariant based on a mapping to a transition between  a bosonic superfluid and a Mott insulator at fractional filling.
By  retaining the same dimer ground states in the presence of further neighbor interactions, namely the six-fold degenerate ground states, the SU(2) invariance of the critical theory should stay intact. In contrast, when the cubic symmetry is suppressed by invoking a symmetry-breaking potential, it has been shown that the transition becomes first-order~\cite{spjb}. 

In order to investigate how the presence of $J_2$ and $J_3$ interactions  affect the nature of the transition, we need to show how the coefficient of the $4$-th order term in the GL functional of this transition changes. Even though the GL functional for such transitions cannot be derived in an exact way, we can use a crude approximation in order to investigate the rough effects of additional interactions. With this in mind, we write firstly the partition function of the system with just $J_1$ interactions present, in terms of Grassmann variables~ \cite{samuel,papanikolaou}:
\begin{eqnarray}
\nonumber
\mathcal{Z} = \int \mathcal{D}\eta \mathcal{D}\eta^\dagger\exp\Big(\frac{1}{2}\sum_{ij}M_{ij}\eta_{i}\eta_{i}^\dagger\eta_{j}\eta_{j}^\dagger \nonumber\\
 + \frac{1}{4}\sum_{ijkl} \tilde J_1M_{ijkl}^{(1)}\eta_{i}\eta_{i}^\dagger\eta_{j}\eta_{j}^\dagger\eta_k\eta_{k}^\dagger\eta_l\eta_{l}^\dagger\Big)\; ,
\end{eqnarray}
where $\tilde J_1=2(e^{J_{1}}-1)$, $M_{ij}$  represents the coordination array of the lattice and $M^{(1)}_{ijkl}$ ($p=1,2,3$) is the interaction matrix which is non-zero and equal to $1$ when the four sites $i$, $j$, $k$, $l$ are the end points of interacting, with strength $J_1$, dimers. 

By using standard methods, we can integrate the Grassmann variables at the expense of introducing two Hubbard-Stratonovich fields and then, by defining the conjugate densities to the fields we can have the Gibbs potential in terms of the densities. Then we write the order parameter of the columnar order as  $m_{\alpha}({\bf r})= (-1)^{r_\alpha} m +m_0$ and we identify one of the two Hubbard-Stratonovich fields by that, while the other Hubbard-Stratonovich field is represented by a spatially uniform function $n$. The Gibbs free energy density in terms of these fields reads:
\begin{eqnarray}
F= 3  \tilde J_1 m^2 + 6 \tilde J_1 m_0^2 + 24 \tilde J_1 m_0 n^2 \nonumber\\ 
+ 3 n^2 + \nonumber \\
\frac{1}{2} ln[36 n^2 (\tilde J_1 (m+2 m_0)+ 1) (\tilde J_1 (-m + 2 m_0) + 1)] \nonumber\\ 
+ \frac{2}{2 \tilde J_1 m_0 +1}.
\end{eqnarray}
which we minimize with respect to $m_0$ and $n$, leaving $m$ as the only order parameter. We need to emphasize here that by minimizing with respect to $m_0$ and $n$, we treat the dimer constraint (which is inherent in the grassmann structure of the theory) in a mean-field fashion. Such an approach is crude, similar in spirit to slave boson approaches for t-J models~\cite{baskaran88}, but it captures basic features of the model, such as tendencies under application of different interactions. The Coulomb-liquid correlations are replaced by isotropic liquid correlations in this case. 

The free energy at the saddle - point can be expanded self-consistently in powers of $m$~\cite{spjb}. Such a mean-field expansion shows a continuous transition at $J_{1}^{c}=0.51$ which we identify as a crude mean-field limit of the transition in the absence of $J_r$ for $r=2$ or $3$. In order to include the additional interactions, we perform a cluster expansion (where the extra terms are proportional to  $(\pm m + m_0)^z$ where $z \geq 3$) to calculate the Gibbs potential in a perturbation expansion in powers of $J_r$ and ultimately, it takes the form~\cite{spjb} $\Gamma(J_1,J_r)/N= C_0(J_1,J_r) + C_2(J_1,J_r)m^2 +C_4(J_1,J_r)m^4 + C_6(J_1, J_r) m^6+... $\; .
 We find that the prefactor $C_4(J_1,J_r)$ is positive for values of  $\tilde J_r$ less than a critical value and becomes negative for larger values. The addition of more terms result in a renormalization of this critical value  where the multicritical point exists to lower values. The accuracy of the method is such that we cannot identify exactly $J_r=0$ as the multicritical point, nevertheless the change of the transition from second to first order can be explained.

%However, the order parameter's susceptibility shows finite-size scaling with a negative anomalous exponent $\eta\simeq-0.45$, which is a clear indication that the transition in this case is also weakly first-order. More details will be presented elsewhere~\cite{spjb}. 

In conclusion, we investigated the phase transition between a columnar crystal and a Coulomb liquid in a 3D dimer system. We showed that attractive perturbations which respect the symmetries of the original Hamiltonian drive the transition first - order. However, when the interactions are repulsive, the transition becomes continuous, seemingly in a different universality class than the one found in Ref.~\cite{alet1}, but a more extensive study is required to characterize appropriately the nature of the transition. In this way, we might have given strong evidence for the nature of the non-compact CP$^1$ universality class.

\begin{acknowledgments}
We thank John Chalker, Claudio Chamon, Paul Fendley, Anders Sandvik, Simon Trebst and Ashvin Vishwanath for stimulating discussions. This research was partially supported by DOE-BES through DE-FG02-07ER46393  (SP).
\end{acknowledgments}

\section{Supplementary material for the manuscript: First- order versus unconventional phase transitions in three-dimensional dimer models}

Here we present evidence that the transition in the repulsive case $J_2<0$ (keeping $J_1>0$),
is continuous, as it appears from the susceptibility and specific heat data. 

 The transition, up to the sizes that we have studied (maximum linear size $L=32$), appears to have similar  characteristics with the transition found by Alet {\it et al.}, but the exponents seem to drift towards different values. We cannot exclude the possibility that the transition is actually in the same universality class, but up to the sizes we studied, such a conclusion is not obvious. The transition has the same characteristics until $J_2$ becomes adequately negative ($J_2\simeq-0.5J_1$), where the 6-fold columnar state becomes unstable towards the formation of $2\times2\times2$ cubes filled with parallel dimers, mutually staggered. This phase is highly degenerate, since the cubes can fluctuate between their different orientations. 
 \begin{figure}[h!]
\includegraphics[width=0.6\columnwidth]{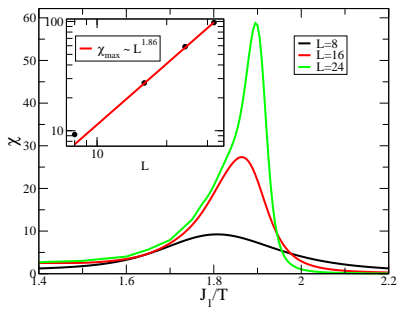}
\centering
\caption{ {\bf Susceptibility of the order parameter $\chi$ and the scaling of its maxima for $J_2/J_1=-0.4$}. The transition is clearly continuous.}
\label{fig:s1}
\end{figure}
 So, for $J_2 < - 0.5J_1$, a novel transition takes place between the Coulomb liquid at high temperatures and a highly degenerate state at low temperatures, characterized by $2\times2\times2$ cubes filled by parallel dimers, which can fluctuate freely. However, this transition is going to be a subject of  a future work. 
 
\begin{figure}[h!]
\includegraphics[width=0.6\columnwidth]{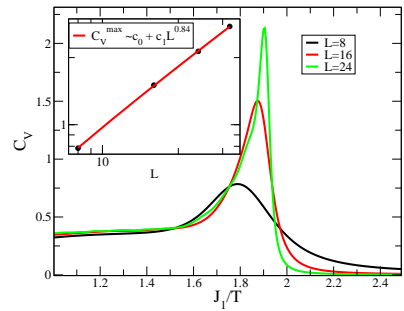}
\centering
\caption{ {\bf The specific heat and the scaling of its maxima for $J_2/J_1=-0.4$}. In this case $\alpha>0$, in contrast with the results of Ref.~\cite{charrier2S}, but close to the values reported in Ref.~\cite{alet05S}.}
\label{fig:s2}
\end{figure}

 In Figs.~\ref{fig:s1}, ~\ref{fig:s2}, we show the behavior of the system in the case $J_2/J_1=-0.4$ which is well deep in the phase which extends for low $J_2$, ranging from $J_2=0$~\cite{alet05S} to $J_2\simeq-0.5J_1$. The results on the critical exponents for smaller $J_2/J_1=-0.1,-0.2$ are consistent with the results we report in Figs.~\ref{fig:s1},~\ref{fig:s2}. In Fig.~\ref{fig:s1} the specific heat shows an instability at $J_1/T_c\simeq1.92$ towards the 6-fold columnar liquid, and up to the sizes studied, the critical exponent for the specific heat, $C_v\sim c_0+c_1 L^{\alpha/\nu}$ is compatible with $\alpha/\nu\simeq 0.75-1.0$. 
 Also, the susceptibility of the columnar order parameter is diverging with the system size according to $\chi\sim L^{\gamma/\nu}\sim L^{2-\eta}$ where $\eta$ is the anomalous exponent of the order parameter. We find, as shown in Fig.~\ref{fig:s2}, that $\gamma/\nu\simeq1.75-2.0$. If we use hyperscaling, namely that $\alpha=2-\nu d$, we find that $\nu=0.5-0.53$. In conclusion, the transition in the repulsive case is continuous and the exponents are close, even though seemingly not the same with the ones that are reported in \cite{alet05S}. The existence of a crossover towards the exponents reported in Ref.~\cite{charrier2S} might be a possibility, but the verification of this scenario requires a more extensive study in much bigger systems.
  \begin{figure}[h!]
\includegraphics[width=0.6\columnwidth]{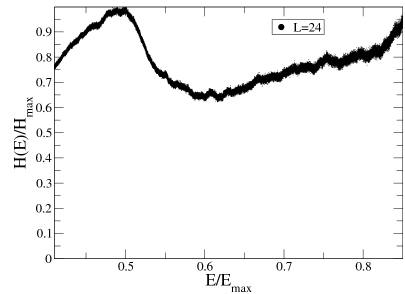}
\centering
\caption{ {\bf Sample histogram for interaction $J_2=0.05J_1$ and system size $L=24$}. For all energies below the maximum possible energy, $E_{\rm max}=J_1(1+J_2/J_1)L^3$ the histogram values are above $80\%$ the average histogram value.}
\label{fig:histogram}
\end{figure}

 The histograms, according to the Wang-Landau recipe, are designed to converge when they are 'flat', closer than $80\%$ of the average histogram value for all recorded energies (being relevant around the transition point)~(cf. Fig.~\ref{fig:histogram}). The procedure of making the histograms flat is repeated for $16$ times, before finally trusting the form of the density of states (again, this is the typical procedure used in the applications of the Wang-Landau algorithm~\cite{wang03S}).

\end{document}